\documentstyle[amssymb,preprint,aps]{revtex}
\begin{document}
\draft
\title{Theory of Nonequilibrium Coherent Transport through \\
an Interacting Mesoscopic Region Weakly Coupled to Electrodes}
\author{{\ Yu Zhu and }Tsung-han Lin$^{*}$}
\address{{\it State Key Laboratory for Mesoscopic Physics and }\\
{\it Department of Physics, Peking University,}{\small \ }{\it Beijing}\\
100871, China}
\author{{\ Qing-feng Sun}}
\address{{\it Center for the Physics of Materials and}\\
{\it Department of Physics, McGill University, Montreal,}\\
{\it PQ, Canada H3A 2T8}}
\date{}
\maketitle

\begin{abstract}
We develop a theory for the nonequilibrium coherent transport through a
mesoscopic region, based on the nonequilibrium Green function technique. The
theory requires the weak coupling between the central mesoscopic region and
the multiple electrodes connected to it, but allows arbitrary hopping and
interaction in the central region. An equation determining the
nonequilibrium distribution in the central interacting region is derived and
plays an important role in the theory. The theory is applied to two special
cases for demonstrations, revealing the novel effects associated with the
combination of phase coherence, Coulomb interaction, and nonequilibrium
distribution.
\end{abstract}

%\vskip 0.4in


\baselineskip 20pt %\baselineskip 12pt
\newpage

In the realm of mesoscopic transport, the basic rules of the traditional
electronics breakdown. When the size of device is smaller than the electron
mean free path in the material, electrons will behave like a wave rather
than particles. Meanwhile, with the reduced size of device, the Coulomb
interaction among electrons becomes important. An additional electron has to
overcome the repulsion from other electrons in the device before entering
it. Moreover, due to lack of inelastic collisions, the thermal distribution
in the device is no longer a equilibrium one when electrodes are biased with
finite voltages. Therefore, a theory containing the above three
``ingredients''---phase coherence, Coulomb interaction, and
nonequilibrium---is of particular interest in the mesoscopic transport.

Using nonequilibrium Keldysh formalism, Meir $et\;al.$ derived a formula, in
which the current flowing out of an electrode is expressed in term of the
Green functions of the central region \cite{current}. The remaining task is
to find out these Green functions. Unfortunately, there seems no standard
method to derive them, although various approximation schemes were used for
special problem in specific system, e.g., large-N expansion \cite{Nexp},
truncation for equation of motion \cite{CBO}, introducing an interpolative
self-energy \cite{Yeyati}, Ng's ansatz for lesser self-energy \cite{Ng},
etc. In fact, too much physics is hidden in the general Hamiltonian, and it
is hopeless to invent a theory covering everything. If we restrict ourselves
to the weak coupling case, the complex Kondo physics will be ruled out. In
these circumstances, the Green functions in the ``atomic limit'' should be a
good starting point for the construction of the full Green functions, and we
shall show that a general solution to the problem is possible. We note that
similar idea has been addressed in the linear response theory by several
authors \cite{dia1,dia2}, but hardly investigated in the nonlinear regime 
\cite{Stafford}.

The aim of this paper is to present a scheme for the calculation of the
Green functions, and hence establish a theory of nonequilibrium coherent
transport through an interacting mesoscopic region. As a price, the coupling
between the central mesoscopic region and the electrodes is required to be
relatively small. Electron transport through the mesoscopic region is viewed
as a summation over various coherent processes via many-body quantum states,
weighted by nonequilibrium thermal probabilities. The many-body quantum
states in the central region can be found by exact diagonalization, while
the nonequilibrium distribution can be determined by an equation derived in
the text .

The mesoscopic system under consideration is modelled by the Hamiltonian
(hereafter $e=\hbar =1$), $H=H_{cent}+\sum_\beta H_\beta +H_T$, where $%
H_{cent}=H_{cent}(\{c_i\},\{c_i^{\dagger }\})$ is a general Hamiltonian for
the central region with $M$-sites (spin index has been absorbed into the
site index $i$), $H_\beta =\sum_\beta (\epsilon _k-V_\beta )a_{\beta
k}^{\dagger }a_{\beta k}$ is the Hamiltonian for the $\beta $th electrodes,
and $H_T=\sum_{\beta ki}\left[ v_{\beta i}a_{\beta k}^{\dagger
}c_i+H.c.\right] $ is the tunnel coupling between them. This Hamiltonian is
applicable to a large variety of mesoscopic systems, such as molecular
devices, tunneling coupled carbon nanotubes, quantum dot arrays,
Aharonov-Bohm rings embedded with quantum dots, etc.

Define the Green function of the central region and the self-energy arise
from the coupling with $\beta $th electrodes as ${\bf G}%
_{ij}^{r,a,<}(t_1,t_2)\equiv \langle \langle c_i(t_1)|c_j^{\dagger
}(t_2)\rangle \rangle ^{r,a,<}$ and ${\bf \Sigma }_{\beta
,ij}^{r,a,<}(t_1,t_2)\equiv \sum_kv_{\beta i}^{*}\langle \langle a_{\beta
k}(t_1)|a_{\beta k}^{\dagger }(t_2)\rangle \rangle _0^{r,a,<}v_{\beta j}$,
where the superscript $r,a,<$ denote for the retarded, advanced, lesser
Green function and self-energy, respectively; the subscript $0$ denotes for
the Green function in the decoupling limit. Following Meir $et\;al.$, the
current flowing out of the $\beta $th electrodes can be expressed in the
compact form $I_\beta =\int \frac{d\omega }{2\pi }Tr\left\{ 2%
%TCIMACRO{\func{Re}}
%BeginExpansion
\mathop{\rm Re}%
%EndExpansion
\left[ {\bf G}(\omega ){\bf \Sigma }_\beta (\omega )\right] ^{<}\right\} $,
where $\left[ AB\right] ^{<}\equiv A^rB^{<}+A^{<}B^a$, ${\bf G}%
^{r,a,<}(\omega )$ and ${\bf \Sigma }^{r,a,<}(\omega )$ are the Fourier
transformed Green function and self-energy. In the wide bandwidth limit, the
self-energy can be evaluated as ${\bf \Sigma }_\beta ^r(\omega )=-\frac{%
\text{i}}2{\bf \Gamma }_\beta $ and ${\bf \Sigma }_\beta ^{<}(\omega )=$i$%
f_\beta (\omega ){\bf \Gamma }_\beta $, where ${\bf \Gamma }_{\beta
,ij}\equiv 2\pi D_\beta v_{\beta i}^{*}v_{\beta j}$ with $D_\beta $ being
the density of states at the Fermi surface of the $\beta $th electrode, $%
f_\beta (\omega )\equiv f(\omega -V_\beta )$ with $f(\omega )$ being the
Fermi distribution function. We assume that the coupling between the central
region and the electrodes is relatively small, i.e., $\Gamma _{\beta ,ij}\ll
k_BT$. Under this assumption, the Green function of the central region can 
{\it approximately} be written in a Dyson equation form ${\bf G=\tilde{g}+%
\tilde{g}\Sigma }_0{\bf G}$,{\bf \ }with ${\bf \tilde{g}\equiv }%
\lim_{H_T\rightarrow 0}{\bf G}$ being the Green function in the decoupling
limit, ${\bf \Sigma }_0\equiv \sum_\beta ({\bf \Sigma }_\beta )$ being the
self-energy contributed by the coupling with electrodes. (This approximation
is amount to a proper truncation of equation of motion.) Accordingly, the
approximate ``Dyson equation'' for ${\bf G}^r$ and the ``Keldysh equation''
for ${\bf G}^{<}$ are 
\begin{eqnarray}
{\bf G}^r &=&{\bf \tilde{g}}^r{\bf +\tilde{g}}^r{\bf \Sigma }_0^r{\bf G}^r%
{\bf \;,} \\
{\bf G}^{<} &=&{\bf G}^r{\bf \Sigma }_0^{<}{\bf G}^a{\bf \;.}
\end{eqnarray}
Consequently, the current formula can be rewritten in a Landauer type as if
in the noninteracting case \cite{current}, 
\begin{equation}
I_\beta =\sum_{\alpha \neq \beta }\int \frac{d\omega }{2\pi }T_{\alpha \beta
}(\omega )\left[ f_\beta (\omega )-f_\alpha (\omega )\right] \;,
\end{equation}
with $T_{\alpha \beta }(\omega )\equiv Tr{\bf G}^r{\bf \Gamma }_\alpha {\bf G%
}^a{\bf \Gamma }_\beta $.

The remaining task is to calculate the retarded Green function in the
decoupling limit. It is straightforward to exactly diagonalize $%
H_{cent}(\{c_i\},\{c_i^{\dagger }\})$ in the particle occupation bases $%
\{(c_M^{\dagger })^{N_M}\cdots (c_2^{\dagger })^{N_2}(c_1^{\dagger
})^{N_1}\left| 0\right\rangle $ where $N_i=0$ or $1$, and obtain the 2$^M$
eigenstates $\{E_n,\left| n\right\rangle \}$. Once again, under the weak
coupling assumption, the density matrix operator of the central region is
supposed to have the {\it diagonal} form ${\bf \rho }_{cent}=\sum_nP_n\left|
n\right\rangle \left\langle n\right| $, with the constraint $\sum_nP_n=1$.
Here, the central region is regarded as ``system'', while the electrodes are
``environment'' in local equilibrium. Given $\{P_n\}$, the decoupled Green
functions can be expressed as 
\begin{eqnarray}
\left\langle \left\langle A|B\right\rangle \right\rangle _0^r &=&\sum_{nm}%
\frac{P_n+P_m}{\omega -(E_n-E_m)+\text{i}0^{+}}\left\langle
m|A|n\right\rangle \left\langle n|B|m\right\rangle \;\;, \\
\left\langle \left\langle A|B\right\rangle \right\rangle _0^{<}
&=&\sum_{nm}2\pi \text{i\ }P_n\delta \left[ \omega -(E_n-E_m)\right]
\left\langle m|A|n\right\rangle \left\langle n|B|m\right\rangle \;\;,
\end{eqnarray}
where $A$ and $B$ are operators composed of $\{c_i\}$ and $\{c_i^{\dagger
}\} $. So the determination of the nonequilibrium distribution $\{P_n\}$
lies in the heart of the theory. In the linear response regime, i.e., $%
\left| V_\beta -V_{\beta ^{\prime }}\right| \ll k_BT$ and hence $V_\beta
\thickapprox V_0$, the central region is in a thermal equilibrium, and the
distribution can be written as $P_n=\frac 1Ze^{-(E_n-N_nV_0)/k_BT}$. For the
case of nonequilibrium, however, $\{P_n\}$ is determined by the coupling to
electrodes with different chemical potentials, and in principle needs a
self-consistent calculation. Our strategy is to choose a proper set of
observables $\{O\}$ and establish the equations of $\{P_n\}$ by the {\it %
stationary} condition 
\begin{equation}
\left\langle \partial _tO\right\rangle =\left\langle \text{i}\left[
H,O\right] \right\rangle =0\;\;.
\end{equation}
We point out that the 2$^M$ conservables $\{O_l\equiv \left| l\right\rangle
\left\langle l\right| \}$ ($\left| l\right\rangle $ is the eigenstate of $%
H_{cent}$) are ideal candidates for the task. Notice that 
\begin{equation}
\left\langle \text{i}\left[ H,O_l\right] \right\rangle =2%
%TCIMACRO{\func{Re} }
%BeginExpansion
\mathop{\rm Re}%
%EndExpansion
\int \frac{d\omega }{2\pi }\sum_{\beta k}\sum_{ij}\left[ \langle \langle
[c_i,O_l]|c_j^{\dagger }\rangle \rangle v_{\beta j}^{*}\langle \langle
a_{\beta k}|a_{\beta k}^{\dagger }\rangle \rangle _0v_{\beta i}\right]
^{<}\;\;,
\end{equation}
and make the approximation $\langle \langle [c_i,O_l]|c_j^{\dagger }\rangle
\rangle \thickapprox \langle \langle [c_i,O_l]|c_j^{\dagger }\rangle \rangle
_0$ under the weak coupling approximation, one can derive a set of equations
of $\{P_n\}$ as 
\begin{equation}
\sum_\beta \sum_{nm}\left[ P_mf_\beta (E_n-E_m)-P_n\bar{f}_\beta
(E_n-E_m)\right] \tilde{\Gamma}_{nm}^\beta Q_{nm}^l=0\;\;\;\;(l=1,2\cdots
2^M)\;,
\end{equation}
where $n$ and $m$ run over all the eigenstates of $H_{cent}$, $\tilde{\Gamma}%
_{nm}^\beta \equiv \sum_{ij}\left\langle m|c_i|n\right\rangle \langle
n|c_j^{\dagger }|m\rangle \Gamma _{\beta ,ij}$, $Q_{nm}^l\equiv \delta
_{nl}-\delta _{ml}$, and $\bar{f}_\beta (\omega )\equiv 1-f_\beta (\omega )$%
. Because $\sum_l\left| l\right\rangle \left\langle l\right| =1$, the 2$^M$
conservables can produce 2$^M$-1 independent equations, and the constraint $%
\sum_nP_n=1$ should be supplemented for completeness. Eq.(8) is the central
result of this work, which determines the nonequilibrium distribution in an
interacting system. With $\{P_n\}$ solved from the set of equations, one can
calculate both nonequilibrium tunneling current and various quantities of
the central region.

To sum up, Eq.(1), (3), (4), and (8) consist of the frame for the
calculation of the nonequilibrium coherent transport through an interacting
mesoscopic region, requiring weak coupling between the central region and
the electrodes, but allowing arbitrary interaction and hopping in the
central region. Below we shall apply the theory to two special cases for
demonstrations.

(1) {\it a single quantum dot with multiple levels. }Consider a single
quantum dot (QD) connected to electrodes with finite bias voltages, which
can be modelled by the Hamiltonian 
\begin{equation}
H_{cent}=\sum_iE_in_i+U\sum_{i<j}n_in_j\;\;,
\end{equation}
where $n_i\equiv c_i^{\dagger }c_i$ is the particle number operator. For
convenience, the particle occupation bases are numbered as $%
F=\sum_{i=1}^MN_i^F\cdot 2^{i-1}$ and $\left| F\right\rangle \equiv
(c_M^{\dagger })^{N_M^F}\cdots (c_2^{\dagger })^{N_2^F}(c_1^{\dagger
})^{N_1^F}\left| 0\right\rangle $. Notice that $H_{cent}$ is already
diagonalized in the $\{\left| F\right\rangle \}$ bases, and the eigenenergy
of $\left| F\right\rangle $ is $E_{\left| F\right\rangle
}=\sum_iE_iN_i^F+U\sum_{i<j}N_i^FN_j^F$. The conservable $\left|
l\right\rangle \left\langle l\right| $ can be written explicitly in the form
of $\{c_i\}$ and $\{c_i^{\dagger }\}$ as $m_M^l\cdots m_2^lm_1^l$ where $%
m_i^l=n_i$ for $N_i^l=1$ and $m_i^l=1-n_i$ for $N_i^l=0$. The equations of $%
\{P_F\}$ are simplified to 
\begin{equation}
\sum_{i=1}^M(-1)^{N_i^l}\Gamma _i\left[ h_i(E_{\left| F_1\right\rangle
}-E_{\left| F_0\right\rangle })P_{F_0}-\bar{h}_i(E_{\left| F_1\right\rangle
}-E_{\left| F_0\right\rangle })P_{F_1}\right] =0\;\;(l=1,2\cdots 2^M)\;,
\end{equation}
where $\Gamma _i\equiv \sum_\beta \Gamma _{\beta i}$, $h_i(\omega )\equiv
\sum_\beta \frac{\Gamma _{\beta i}}{\Gamma _i}f_\beta (\omega )$, $\Gamma
_{\beta i}\equiv \Gamma _{\beta ,ii}=2\pi D_\beta |v_{\beta i}|^2$, $%
F_1=l-N_i^l\cdot 2^{i-1}+2^{i-1}$, $F_0=l-N_i^l\cdot 2^{i-1}$, and $\bar{h}%
_i(\omega )\equiv 1-h_i(\omega )$. The retarded Green function ${\bf \tilde{g%
}}^r(\omega )$ is obtained as 
\begin{equation}
\langle \langle c_i|c_j^{\dagger }\rangle \rangle _0^r=\delta _{ij}\sum_F%
\frac{P_F}{\omega -\tilde{E}_i^F+\text{i}0^{+}}\;\;,
\end{equation}
with $\tilde{E}_i^F\equiv E_i+U\sum_{j\neq i}N_j^F$ being the renormalized
resonances. Specially, for $M=2$, using $\langle n_1\rangle =P_{01}+P_{11}$, 
$\langle n_2\rangle =P_{10}+P_{11}$, $\langle n_1n_2\rangle =P_{11}$,
Eq.(19) and Eq.(20) are equivalent to 
\begin{eqnarray}
\langle n_i\rangle &=&(1-\langle n_{\bar{i}}\rangle )h_i(E_i)+\langle n_{%
\bar{i}}\rangle h_i(E_i+U)\;\;, \\
\langle n_1n_2\rangle &=&\frac{\Gamma _1}{\Gamma _1+\Gamma _2}%
h_1(E_1+U)\langle n_2\rangle +\frac{\Gamma _2}{\Gamma _1+\Gamma _2}%
h_2(E_2+U)\langle n_1\rangle \;\;, \\
\langle \langle c_i|c_i^{\dagger }\rangle \rangle _0^r &=&\frac{1-\langle n_{%
\bar{i}}\rangle }{\omega -E_i+\text{i}0^{+}}\;+\frac{\langle n_{\bar{i}%
}\rangle }{\omega -E_i-U+\text{i}0^{+}}\;,
\end{eqnarray}
with $\bar{i}=3-i$ for $i=1$ or $2$. Thus, our theory reproduces the correct
results for the occupation number $\langle n_i\rangle $ and the retarded
Green function $\langle \langle c_i|c_i^{\dagger }\rangle \rangle _0^r$ in
the limit of $\Gamma _i\rightarrow 0$ \cite{CBO}, and derive the correlator $%
\langle n_1n_2\rangle $ which is otherwise difficult to obtain. Fig.1 shows
the equilibrium and nonequilibrium distributions for the quantum dot
containing two interacting levels connected with two electrodes. One can see
in the plot: (a) The complete Coulomb blockade in equilibrium is partially
removed in nonequilibrium, i.e., the blockaded state can be occupied in some
``windows'' of the gate voltage. (b) The correlator $\langle n_1n_2\rangle $
is obviously unequal to $\langle n_1\rangle \langle n_2\rangle $ in
nonequilibrium, although approximately correct in equilibrium for
nondegenerate levels. (c) The total occupation number has fractional steps
in nonequilibrium in contrast to integer steps in equilibrium, and the
fluctuation of the total number is reminiscent of the shape of tunneling
current.

Next, we insert the QD to one arm of a Aharonov-Bohm (AB) interferometer 
\cite{Yacobi}. The other arm (reference arm) is modelled by a quantum point
contact, which can be described by adding the term $\sum_k(We^{\text{i}\phi
}a_{Lk}^{\dagger }a_{Rk}+H.c.)$ to $H_T$ \cite{Fano}, with $\phi $ being the
AB phase induced by magnetic flux. Both the current formula and the equation
of $\{P_n\}$ should be modified to include the ``direct'' coupling between
electrodes, the details will be presented elsewhere. The background
conductance of the reference arm (measured when QD is decoupled from both
electrodes) is $G_0=\frac{e^2}h\frac{4x}{(1+x)^2}$ with $x\equiv \pi
^2W^2D_LD_R$. The effective conductance of QD is formally defined as $%
G_{dot}=I(V_b,V_g)/V_b-G_0$. Fig.2 shows the curves of $G_{dot}$ in both
linear and nonlinear transport regimes. Three features are noteworthy: (a)
In linear transport, the conductance is contributed only by the ground
state, while in nonlinear transport, the conductance is contributed by both
ground and excited states, recognized by the substeps in the conductance
plateaus and valleys, which is in agreement with the recent experiment \cite
{neqb}. (b) With the increase of the background conductance, $G_{dot}$ vs $%
V_g$ curves exhibit Fano pattern, evolving from a peak (plateau) to a pair
of peak and dip (plateau and valley), and finally to a dip (valley), which
is originated from the constructive and destructive interference between a
resonance and the uniform background. (c) The phase analysis shows that the
dependence of $G_{dot}$ on $\phi $ across a resonance is quite different
between linear and nonlinear transport. In linear transport, the dependence
changes from $\cos \phi $ to $\cos (\phi +\pi )$ abruptly across a
resonance, and $\cos (2\phi )$ component only appears on the resonance. In
nonlinear transport, $\cos (2\phi )$ component is always accompanied with
the $\cos \phi $ component, and the crossover from $\cos \phi $ to $\cos
(\phi +\pi )$ occurs continuously. The ``nonequilibrium-Fano'' effect
discussed here and the ``Kondo-Fano'' effect in \cite{Fano} seem to share
some similarities, although the mechanisms are basically different.

(2) {\it coupled double quantum dots.} Consider two quantum dots coupled in
series with left and right electrodes (N-QD=QD-N), each dot is capacitively
coupled to a gate so that the energy level of the dot is tunable \cite{CQD}.
The coupled double quantum dots can be modelled by a 4-site Hamiltonian 
\begin{eqnarray}
H_{cent} &=&\sum_{i=1,2}\sum_\sigma E_{i\sigma }n_{i\sigma }+t\sum_\sigma
(c_{1\sigma }^{\dagger }c_{2\sigma }+c_{2\sigma }^{\dagger }c_{1\sigma }) 
\nonumber \\
&&+\sum_{i=1,2}U_in_{i_{\uparrow }}n_{i_{\downarrow }}+U_{12}(n_{1\uparrow
}+n_{1\downarrow })(n_{2\uparrow }+n_{2\downarrow })\;\;.
\end{eqnarray}
We first diagonalize $H_{cent}$ in the 2$^4$ particle occupation bases. Due
to the particle number conservation and spin conservation, the 16
dimensional spaces can be divided into several subspaces $%
16=1+(2+2)+(1+1+4)+(2+2)+1$, in which eigenstates are readily solved. In
principle, one can find out 2$^4$ conservables written in $\{c_{i\sigma }\}$
and $\{c_{i\sigma }^{\dagger }\}$ as done in (1), although it is uneasy and
unnecessary. We only need to calculate the effective coupling strength $%
\tilde{\Gamma}_{nm}^\beta $ and put them into Eq.(8). With $\{P_n\}$ solved
from the 2$^4$ linear equations, ${\bf \tilde{g}}^r$, hence ${\bf G}^r$ and $%
{\bf G}^{<}$ are available. Finally, the current formula in the N-QD=QD-N
system can be simplified as 
\begin{equation}
I\equiv I_L=-I_R=\sum_\sigma \int \frac{d\omega }{2\pi }\Gamma _L\Gamma
_R\left| \langle \langle c_{1\sigma }|c_{2\sigma }^{\dagger }\rangle \rangle
^r\right| ^2\left[ f_L(\omega )-f_R(\omega )\right] \;\;.
\end{equation}
We present the numerical results of the tunneling current $I$ vs the
resonant levels $(E_1,E_2)$ in Fig.3. The bias voltage $V_b$ ($V_L=V_b/2$, $%
V_R=-V_b/2$) is fixed as $0.001$, $0.3$, and $0.6$ for the graphs from top
to bottom. The graph of $V_b=0.001$ is corresponding to the linear response
regime, in which the thermal distribution in the central region is nearly
equilibrium. Due to the intradot and interdot Coulomb interactions, the
conductance peaks are arranged to a hexagon pattern, consistent with both
experiment and the semiclassical Coulomb blockade model. The graph of $%
V_b=0.6$ is corresponding to the strong nonequilibrium case, in which
thermal distribution of the central region is determined by two reservoirs
with different chemical potentials. The effect of finite bias voltage on the
tunneling current is two folded: (a) Pull each conductance peak along the
direction of $E_1=E_2$, and ultimately emerge them into a ridge. It is easy
to understand the pulling effect by thinking of the noninteracting case
(i.e., $U_1=U_2=U_{12}=0$), where the conductance through N-QD=QD-N is
allowed only when $V_L>E_1\thickapprox E_2>V_R$. (b) Modify the hexagon
pattern significantly, and break the symmetry with respect to $E_1=E_2$.
This can be attributed to the nonequilibrium occupation of the coupled
quantum dots. To proceed, let us cut off the interdot hopping for a moment
(i.e., $t=0$). Then the electron filling to the left (right) dot is only
related to the chemical potential of the left (right) electrode. The
occupation configuration $\{\langle n_1\rangle ,\langle n_2\rangle \}$ vs
the energy levels $(E_1,E_2)$ has the same shape of hexagon boundary as in
the equilibrium case, except for a displacement of $(\frac{V_b}2,-\frac{V_b}2%
)$. Therefore, the symmetry with respect to $E_1=E_2$ is broken when $%
V_b\neq 0$. Turning on the interdot hopping will make the problem much more
complicated, either $\langle n_1\rangle $ or $\langle n_2\rangle $ are not
good quantum number, energy levels of each dot are hybridized into
``molecular orbits'', and the occupation of the ``molecule'' is affected by
both of the electrodes. Simple interpretation for this situation is beyond
our intelligence.

In conclusion, we have presented a theory dealing with nonequilibrium
coherent transport through an arbitrary mesoscopic region, possibly
containing strong Coulomb interactions. The only restriction of the theory
is that the coupling between the central region and electrodes should be
sufficient weak so that the central region may be regarded as a single
quantum system. The key innovation of the theory is Eq.(8) which determines
the nonequilibrium distribution in an interacting system. The general theory
is applied to two special cases, a single quantum dot with multiple
interacting levels and coupled double quantum dots, and novel behaviors are
found in the nonlinear coherent transport.

We would like to thank W. Li and Y. F. Yang for the valuable discussions.
This project was supported by NSFC\ under Grant No. 10074001 and by the
State Key Laboratory for Mesoscopic Physics in Peking University.

\smallskip $^{*}$ To whom correspondence should be addressed.

%\section* {REFERENCES}

\newpage

\section*{Figure Captions}

\begin{itemize}
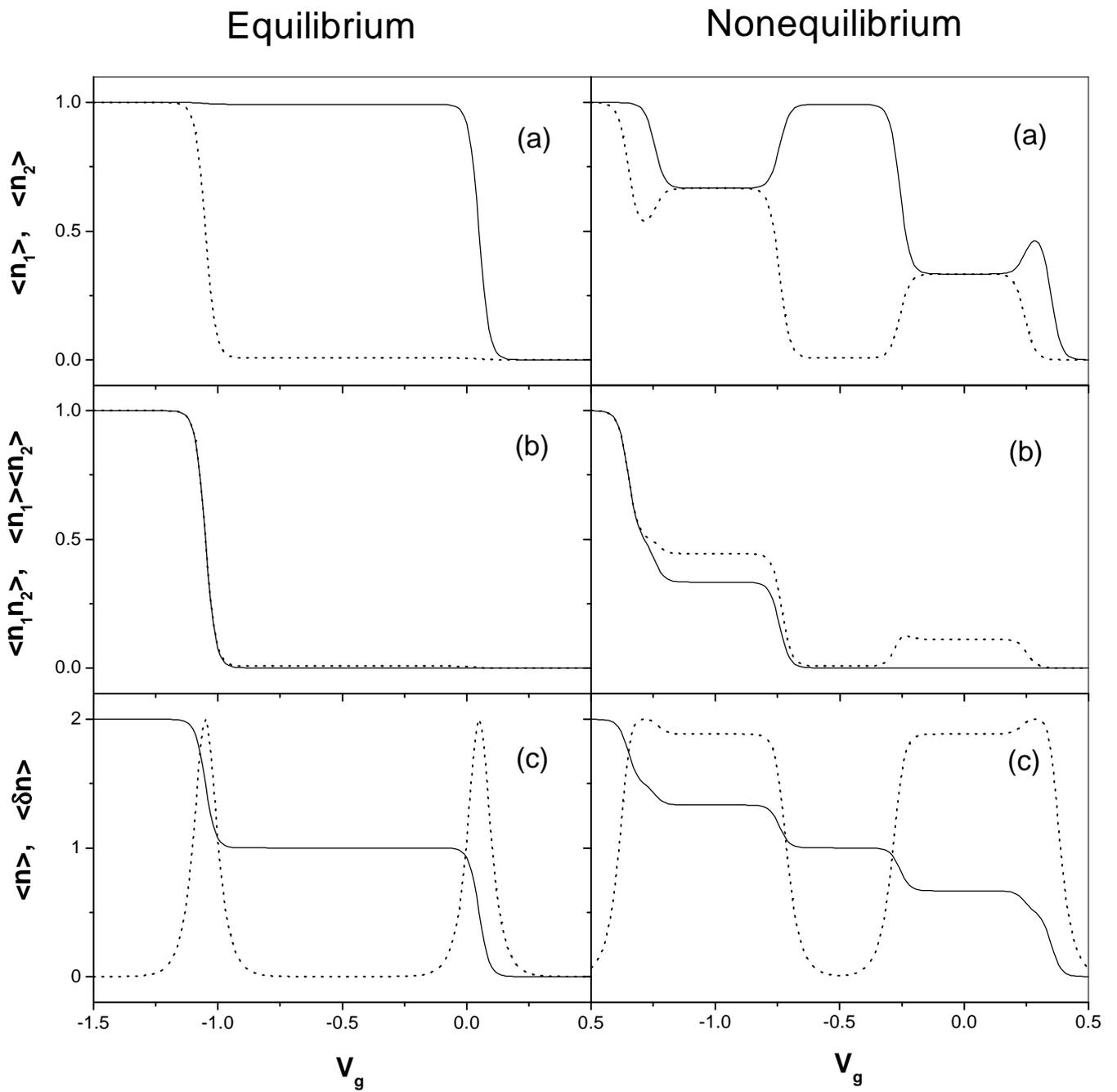
\item[{\bf Fig. 1}]  Equilibrium and nonequilibrium distributions in a QD
containing two energy levels indexed by $i=1,2$, connected with two
electrodes indexed by $\beta =L,R$. (a), (b) and (c) show the curves of $%
\langle n_1\rangle $ (solid) and $\langle n_2\rangle $ (dotted), $\langle
n_1n_2\rangle $ (solid) and $\langle n_1\rangle \langle n_2\rangle $
(dotted), $\langle n\rangle \equiv \langle n_1\rangle +\langle n_2\rangle $
(solid) and $\langle \delta n\rangle \equiv 4\left[ \langle n^2\rangle
-\langle n\rangle ^2\right] ^{1/2}$ (dotted) vs the gate voltage $V_g$,
respectively. The bias voltage $V_b\equiv V_L-V_R$ is fixed at $V_b=0$ / $%
V_b=0.6$ for the left / right panel. Other parameters are: $U=1$, $k_BT=0.02$%
, $\Gamma _{1L}=\Gamma _{1R}=\Gamma _{2L}=\Gamma _{2R}=0.001$, $E_1=V_g-0.05$%
, $E_2=V_g+0.05$.

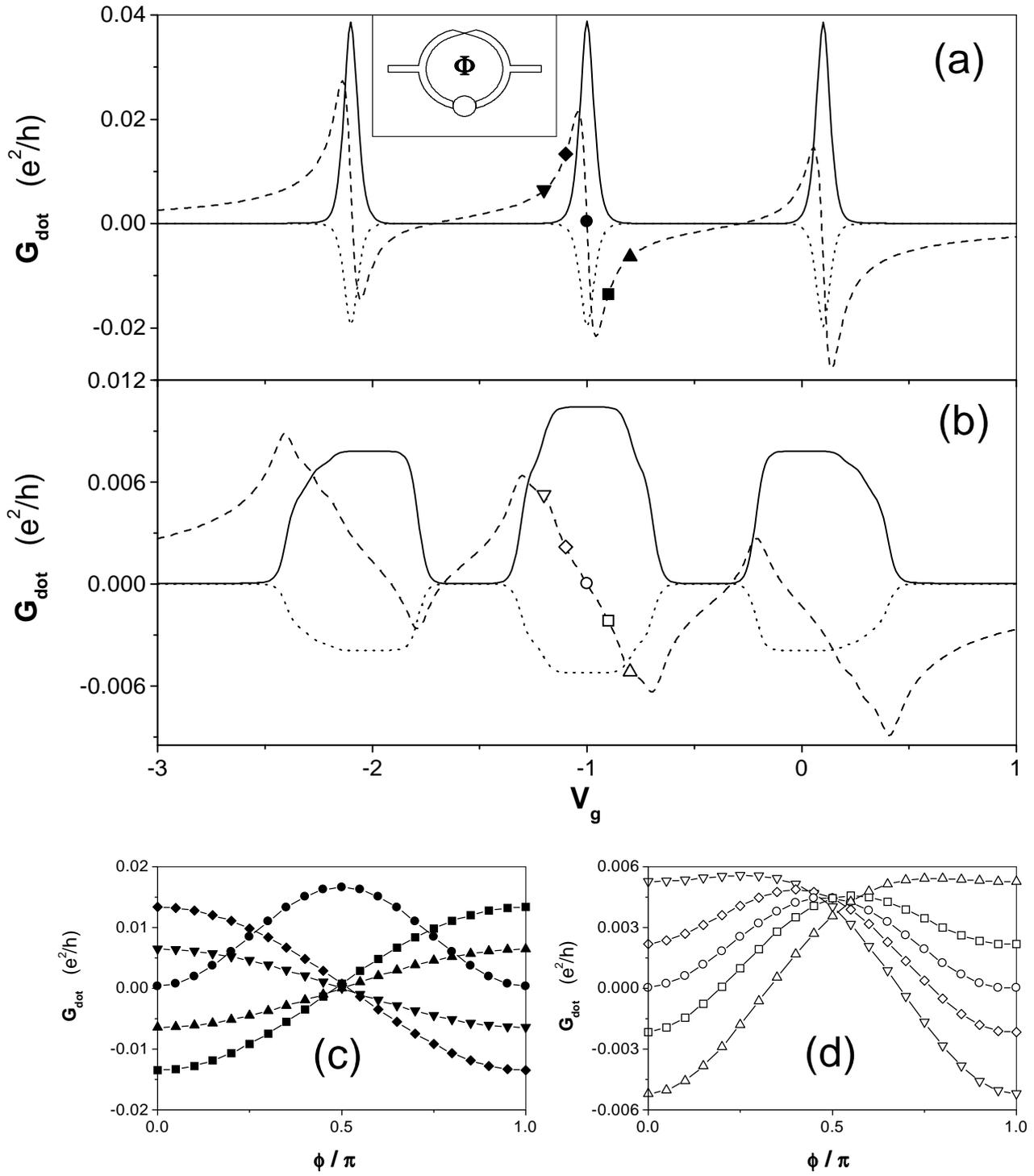
\item[{\bf Fig. 2}]  Linear and nonlinear transport through an AB
interferometer embedded with a QD (schematically shown in the inset). (a)
and (b) show the curves of $G_{dot}\equiv I(V_b,V_g)/V_b-G_0$ vs the gate
voltage $V_g$ for $\phi =0$, with the bias voltage $V_b$ fixed at $0^{+}$
and $0.6$, respectively. The solid, dash, and dotted curves correspond to
the background conductance of the reference arm $G_0=0$, $0.5\frac{e^2}h$, $%
\frac{e^2}h$, respectively. The QD contains three interacting levels, with
level spacing $\Delta E=0.1$, interacting constant $U=1$, and $\Gamma
_L=\Gamma _R=0.001\ll k_BT=0.02$. (c) and (d) analyze the phase dependence
of the $G_{dot}$ at the points marked in (a) and (b). Only the range of $%
0<\phi <\pi $ is shown since $G_{dot}(\phi )=G_{dot}(-\phi )$.

\item[{\bf Fig. 3}]  Surface graphs of tunneling current $I$ vs the resonant
levels $(E_1,E_2)$ in the N-QD=QD-N system, with sideview on the left and
topview on the right. The bias voltage is fixed at $V_b=0.001$,$\;0.3$,$\;$%
and $0.6$ for the graphs from top to bottom. Other parameters are: $U_1=U_2=1
$, $U_{12}=0.3$, $t=0.1$, $k_BT=0.05$, $\Gamma _L=\Gamma _R=0.01$. (In the
e-print version, Fig. 3 is separated into Fig. 3a, Fig.3b and Fig. 3c
corresponding to $V_b=0.001$,$\;0.3\;$and $0.6$ respectively,  to reduce the
size of figures.)
\end{itemize}


\begin{references}
\bibitem{current}  Y. Meir and N. S. Wingreen, Phys. Rev. Lett. {\bf 68,}
2512 (1992).

\bibitem{Nexp}  N. E. Bickers, Rev. Mod. Phys. {\bf 59,} 845 (1987).

\bibitem{CBO}  Y. Meir, N. S. Wingreen, and P. A. Lee, Phys. Rev. Lett. {\bf %
66,} 3048 (1991).

\bibitem{Yeyati}  A. L. Yeyati, F. Flores, and A. Martin-Rodero, Phys. Rev.
Lett. {\bf 83,} 600 (1999).

\bibitem{Ng}  T. -K. Ng, Phys. Rev. Lett. {\bf 76,} 487 (1996).]

\bibitem{dia1}  J. M. Kinaret $et\;al.$, Phys. Rev. B {\bf 45, }9489 (1992).

\bibitem{dia2}  G. Chen, G. Klimeck, and S. Datta Phys. Rev. B {\bf 50,}
8035 (1994).

\bibitem{Stafford}  We are only aware of the paper by C. A. Stafford, Phys.
Rev. Lett. {\bf 77,} 2770 (1996), which is devoted to the identical problem.
However, his theory requires the ground state to be nondegenerate, the
temperature and the bias voltages are small compared to the excitation
energy, while our theory release these restrictions, and is a {\it real}
nonequilibrium transport theory.

\bibitem{Yacobi}  A. Yacoby $et\;al.$, Phys. Rev. Lett. {\bf 74,} 4047
(1995).

\bibitem{Fano}  W. Hofstetter, J. K\"{o}nig, and H. Schoeller, Phys. Rev.
Lett. {\bf 87,} 156803 (2001).

\bibitem{neqb}  M. M. Deshmukh $et\;al.$, Phys. Rev. B {\bf 65, }073301
(2002).

\bibitem{CQD}  F. R. Waugh $et\;al.$, Phys. Rev. Lett. {\bf 75, }705 (1995);
C. Livermore $et\;al.$, Science {\bf 274, }1332 (1996).
\end{references}
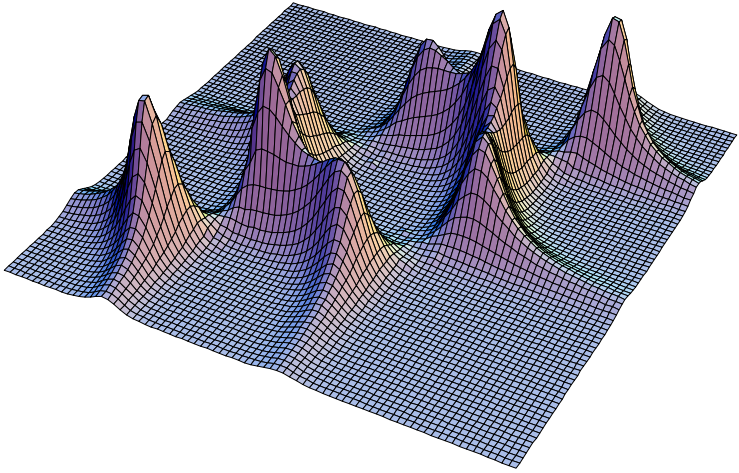
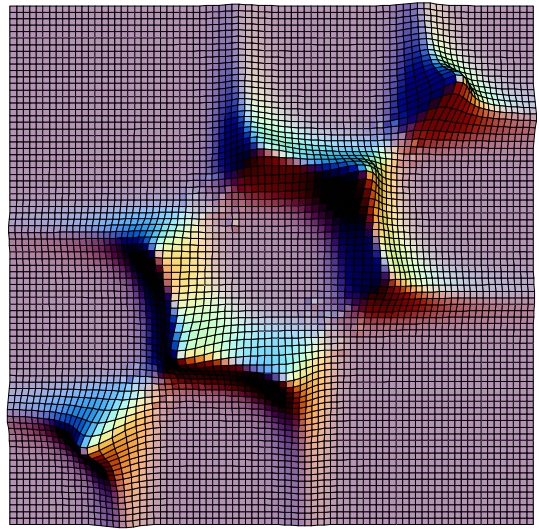
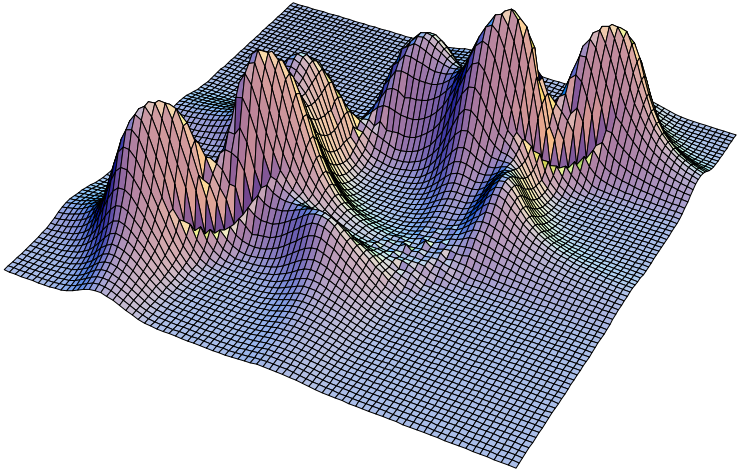
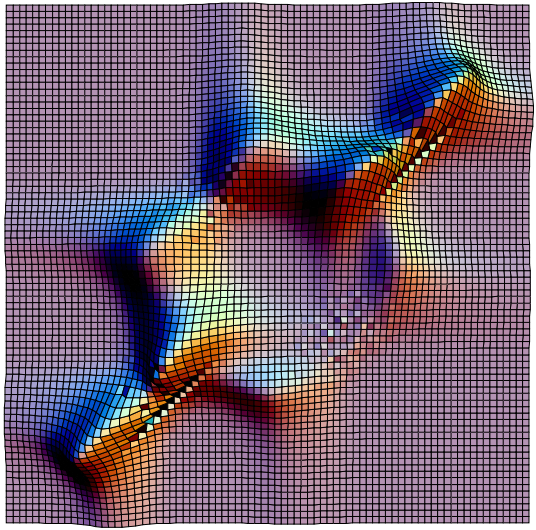
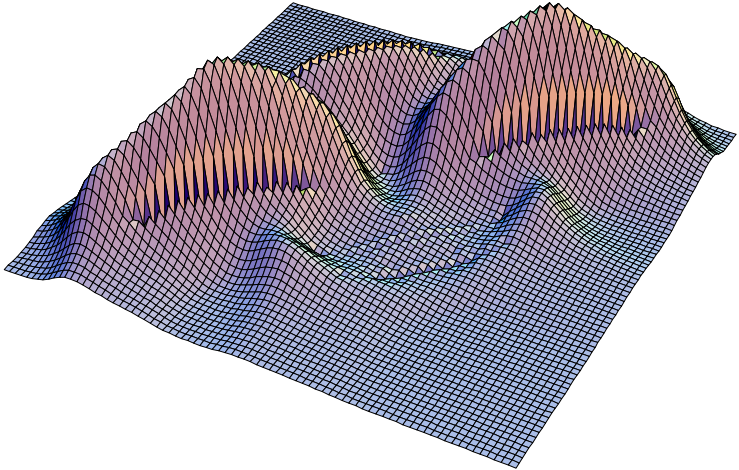
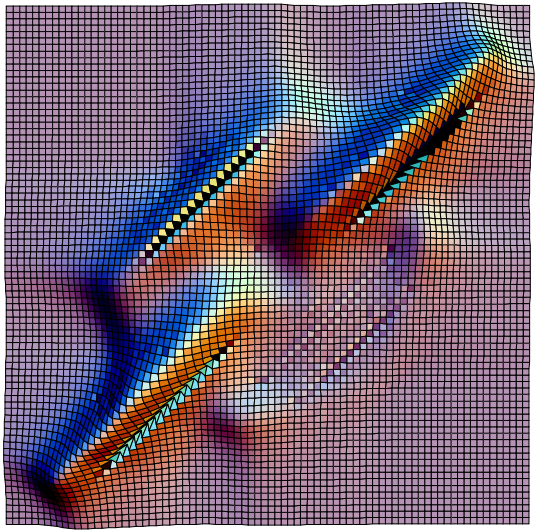
\end{document}